\documentclass{ifacconf}
\usepackage{lmodern}
\usepackage[T1]{fontenc}
\usepackage{lmodern}
\usepackage{textcomp}
\usepackage{graphicx}      % include this line if your document contains figures
\usepackage{natbib}
\usepackage{amsmath} 
\usepackage{bbm}% required for bibliography
\usepackage{bbm}

\usepackage{algorithm}
\usepackage{algpseudocode}
\makeatother
\begin{document}
\begin{frontmatter}
\vspace{-20pt}

\makebox[0pt][c]{\raisebox{1pt}{%

  \parbox{\textwidth}{\centering

    \small © 2026. This work has been accepted to IFAC for publication under a Creative Commons Licence CC-BY-NC-ND and will be presented at the Modeling, Estimation and Control Conference (MECC 2026) in Phoenix, Arizona, USA.}
}}
 
\vspace{-15pt}
 
\title{3D Modeling of a Tethered Autogyro with Articulated Rotors and Attitude Control using Differential Rotor Braking} 

\author[First]{Tasnia Noboni} 
\author[First]{Tuhin Das} 

\address[First]{Department of Mech. and Aero. Eng., Univ. of Central Florida, FL 32816, USA (E-mail: Tasnia.Noboni@ucf.edu, Tuhin.Das@ucf.edu).}
%\address[Second]{Department of Mechanical and Aerospace Engineering, University of
%Central Florida, FL 32816, USA (e-mail: Tuhin.Das@ucf.edu)}

\begin{abstract}     % Abstract of 50--100 words
A tethered autogyro with articulated rotors can operate as an unmanned aerial vehicle capable of energy-efficient, long-duration deployment by utilizing ambient wind energy to sustain flight. This article presents a model-based attitude control technique for such a system using a full-fidelity dynamic model. Using Lagrangian approach combined with Blade Element Momentum Theory and catenary mechanics, a previously developed 2D hybrid model is extended to three dimensions. The new model describes the complete rigid-body motion of the frame, including roll and yaw dynamics, and is augmented with rotor speed and flapping degrees of freedom for each blade. Equilibrium characteristics are examined through steady-state responses and compared with the prior model. The resulting trends of equilibria are found to be consistent with those reported in the literature. A feedback control strategy based on regenerative differential rotor braking is developed to modulate all three attitude angles. Simulations demonstrate effective attitude regulation and stable flight. 
\end{abstract}


\end{frontmatter}
%===============================================================================
\vspace{-0.18in}
\section{Introduction}
\vspace{-0.12in}
%Autogyro produces lift through autorotation of unpowered rotors in a strong wind field. When tethered to the ground, it can serve as an energy-efficient platform for long-duration monitoring without requiring continuous external power, as it uses ambient wind energy to stay aloft. Therefore, investigating such tethered systems is important for practical airborne monitoring applications.
A tethered autogyro relies on wind-driven autorotation of its unpowered rotors to stay aloft without continuous external power, making it a promising energy-efficient platform for long-duration airborne monitoring. Understanding the dynamics of such tethered systems is therefore essential for practical airborne monitoring applications and effective control design. 
Although extensive modeling efforts exist for wind-energy extraction systems, including offshore wind turbine structural and hydrodynamic dynamics \citep{sakif2026improved, rahman2025modeling}, comparatively little attention has been given to the modeling and control of autogyro-based airborne systems.

The aerodynamic modeling of autogyros using the Blade Element Momentum Theory (BEMT) has evolved over time. Early studies assumed constant blade pitch \citep{Glauert26}, while later work incorporated linearly varying pitch distributions with experimental validation \citep{wheatley1935aerodynamic}. Unlike conventional helicopters with nearly constant rotor speed, autogyros exhibit variable rotor speed, requiring its inclusion as an additional degree of freedom in extended rotorcraft models \citep{lopez2004dynamics,thomson2005application}. Building on earlier aerodynamic analyses \cite{wheatley1935aerodynamic}, a more recent study examined the steady-state equilibrium characteristics using a 2D model of tethered autogyro~\citep{mcconnell2022equilibrium}. However, detailed dynamic modeling and control of tethered autogyros remain limited, with only a few works addressing specific stability and control aspects \citep{noboni2024altitude, noboni2025,noboni2025adaptive}.

The tether further complicates the dynamics by coupling rigid-body motions with the tether, which affects maneuverability and stability. Earlier studies analyzed longitudinal stability using linearized models with simplified massless cable \citep{rye1985longitudinal}, and configurations with teetering rotors \citep{houston1998identification}. In our previous study \citep{10155811}, regenerative differential rotor braking was used to regulate pitch and altitude using a reduced-order autogyro model. These results showed that equilibrium altitude depends on pitch angle, increasing up to an optimal value and decreasing beyond it. This trend was later confirmed using a higher-order dynamic model in \cite{noboni2025}, which adopted a hybrid modeling framework between 2D and 3D, included blade-flapping degrees of freedom, and relaxed the thrust-direction assumptions.

In this work, we extend the modeling framework in \cite{noboni2025} to a full three-dimensional tethered multirotor autogyro model by relaxing the vertical planar-motion constraint and incorporating full rigid-body dynamics. Based on this control-oriented model, a feedback attitude controller using regenerative differential rotor braking is developed, and the equilibrium characteristics are compared with those of the 2D model.
\vspace{-0.1in}
\section{Modeling Approach}
\begin{figure}[htpb]
\begin{center}
\includegraphics[width=0.40\textwidth]{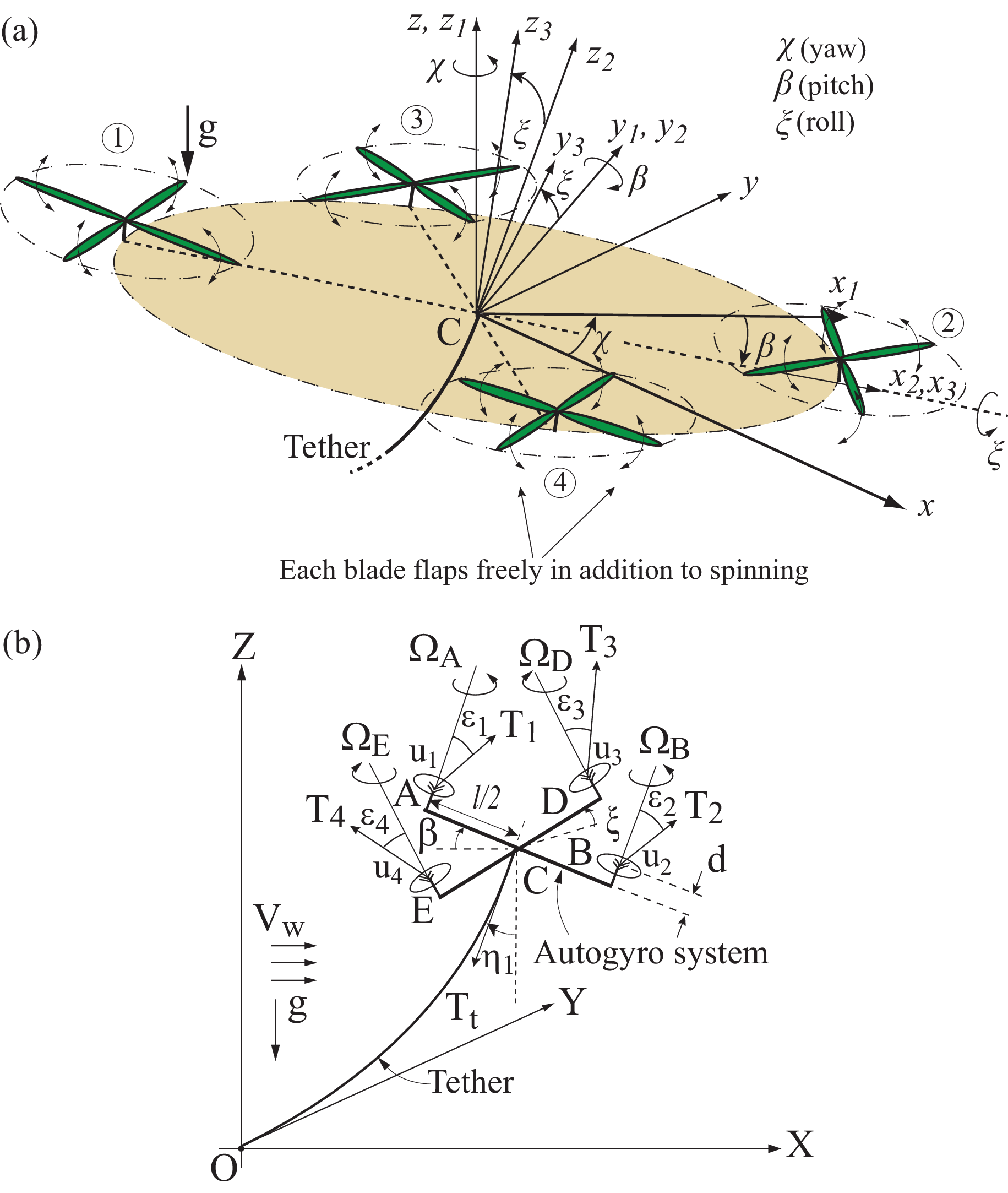}
	\caption{Tethered autogyro system: (a) Euler angles; (b) Quadrotor system}
	\label{fig:sys}
\end{center}
%\vspace{-0.1in}
\end{figure}

\vspace{-0.15in}
\label{model_sd}
A quadrotor configuration with four equispaced blades per rotor, shown in Fig.~\ref{fig:sys}(a), is adopted for the autogyro studied in this paper. This work extends the modeling framework in \cite{noboni2025} to a full three-dimensional formulation by incorporating roll and yaw dynamics. In the earlier model, the system motion was constrained to the X–Z plane, and the roll and yaw motions were assumed to be regulated by the lateral rotors. These assumptions are relaxed here by explicitly modeling the lateral rotor dynamics.
%A quadrotor configuration with four equispaced blades per rotor, as shown in Fig.~\ref{fig:sys}(a), is adopted for the autogyro studied in this paper. This work builds upon the dynamic model presented in \cite{noboni2025} by incorporating the yaw and roll dynamics of the quadcopter. The prior work in \cite{noboni2025} developed a 2D version of Fig.~\ref{fig:sys} detailing 3D aerodynamic forces and blade-level dynamics, with blade flapping modeled as an additional degree of freedom. The roll and yaw motions were assumed to be regulated by the lateral rotors, restricting system motion to the X–Z plane. In this paper, these assumptions are relaxed by explicitly modeling lateral rotor dynamics to obtain a comprehensive three-dimensional formulation. 
The resulting Lagrangian-based model provides a higher-fidelity representation of the system dynamics. The corresponding 3D configuration and parameters are shown in Fig.~\ref{fig:sys}(b).

\begin{figure} [htbp]
\begin{center}
\includegraphics[width=0.45\textwidth]{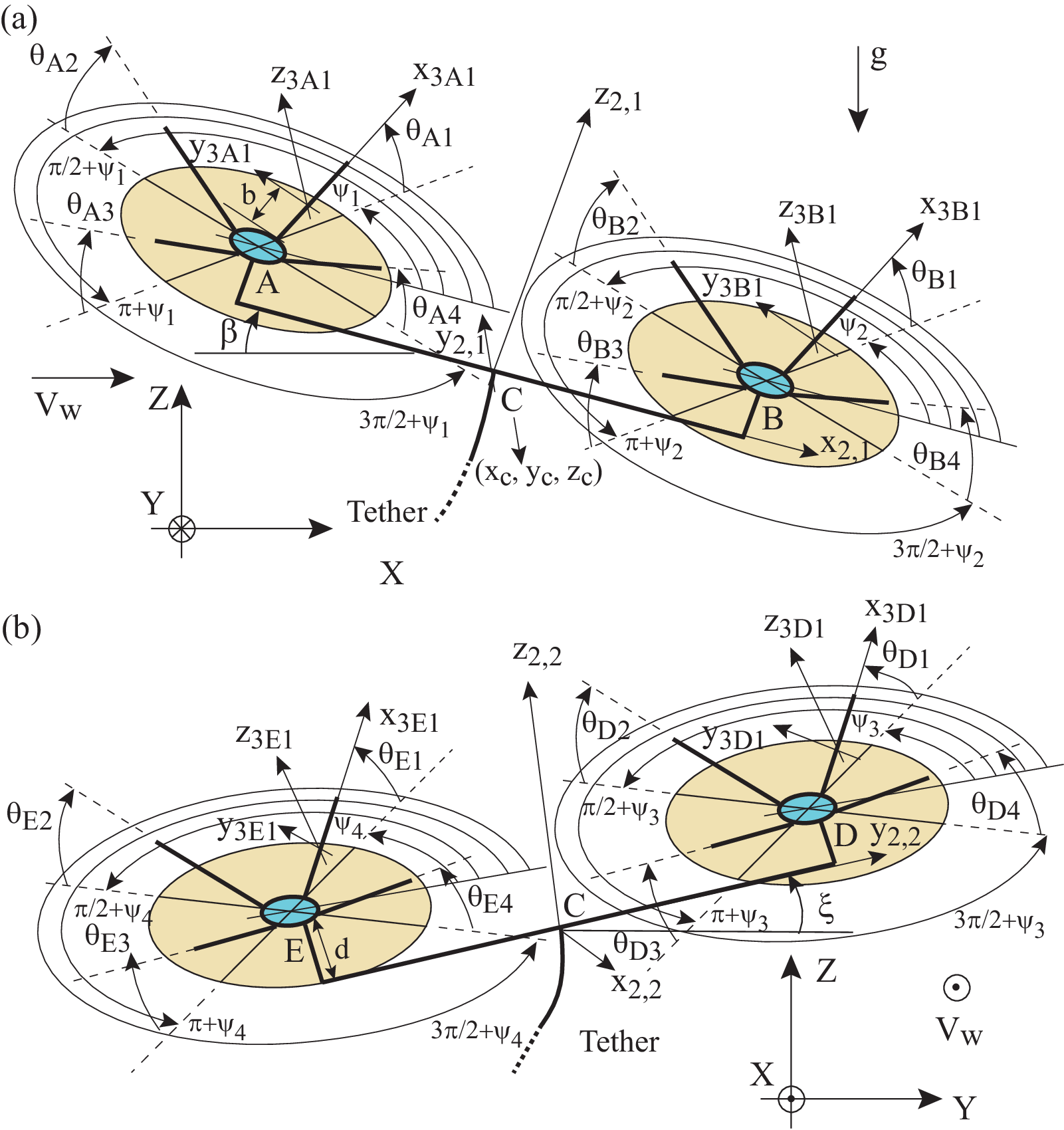}
	\caption{Reference frames and the generalized coordinates of the tethered autogyro in 2D plane (a) longitudinal rotors in X-Z plane; (b) lateral rotors in Y-Z plane}
	\label{fig:gc}
\end{center}
\end{figure}

Figure~\ref{fig:gc} illustrates the reference frames and generalized coordinates. The flapping degree of freedom for each blade is represented by $\theta_{\mathcal{R}j}, \,\mathcal{R} \in \{A, B, D, E\},\, j=1,\ldots,4$. The angles $\psi_i$ represent the rotation of the hubs, which lie in the $x_{2}$-$y_{2}$ plane. Here, $i=1,\ldots,4$ refer to rotors 1,2,3, and 4, centered at $A$, $B$, $D$, and $E$  respectively. In Fig.~\ref{fig:gc}, $C$ is located at $(x_c, \,y_c,\,z_c)$ and the frame has a pitch inclination of $\beta$. 
The full 3D model includes yaw and rolling DoFs. This results in an increase in the number of states from 26 in \cite{noboni2025} to 52 in this paper. The generalized coordinates are,
\small
\begin{equation}
\begin{aligned}
\mathbf{q} = [&x_c\; y_c\; z_c\; \xi\; \beta\; \chi\; \psi_1\;
\theta_{A1}\; \theta_{A2}\; \theta_{A3}\; \theta_{A4}\;\psi_2\;\theta_{B1}\; \theta_{B2}\; \theta_{B3}\; \theta_{B4}\; \\
& \psi_3\; \theta_{D1}\; \theta_{D2}\; \theta_{D3}\; \theta_{D4}\; \psi_4 \; \theta_{E1}\; \theta_{E2}\; \theta_{E3}\; \theta_{E4}]^{T}
\end{aligned}
\label{gen_coord}
\end{equation}
\normalsize
\noindent where, $\theta_{\mathcal{R}j}, \,\mathcal{R} \in \{A, B, D, E\},\, j=1,\dots,4$, denotes flapping angle of each blade. The three types of coordinate systems used to describe the system dynamics as shown in Fig.~\ref{fig:gc}, are the inertial reference frame $(X,Y,Z)$, the system fixed to the frame at $C$, ($x_{2,i},y_{2,i},z_{2,i}$), and the coordinate systems fixed to the center of mass of each blade, namely, ($x_{3\mathcal{R}j},y_{3\mathcal{R}j},z_{3\mathcal{R}j}$). Thus, a $z-y-x-z-y$ Euler angle rotation sequence is utilized to obtain the orientation of each blade consisting of: 1) Rotation by $\chi$ about $Z$ axis, 2) Rotation by $\beta$ about $Y$ axis, 3) Rotation by $\xi$ about $X$ axis, 4) Rotation by $(\psi_i+n\frac{\pi}{2})$ about $z_{3}$ direction, 5) Rotation by $-\theta_{\mathcal{R}j}$ about $y_{3\mathcal{R}j}$, $\mathcal{R} \in \{A, B, D, E\},\, j = 1,\dots 4$. Here, $i=1\dots4$ represents rotor number and $n$ will vary from 0 to 3 for each $i$, indicating the blades. The rotation matrices are,
{\footnotesize
\begin{equation}
    \begin{aligned}
    &\mathbf{R}_{\xi,x}=\begin{bmatrix}
    1&0&0\\
	0&\cos\xi&\sin\xi\\
	0&-\sin\xi&\cos\xi\\
    \end{bmatrix}\!\!, \;\mathbf{R}_{\beta,y}=
    \begin{bmatrix}
	\cos\beta&0&-\sin\beta\\
	0&1&0\\
	\sin\beta&0&\cos\beta\\
    \end{bmatrix}\!\!,\\
    &\mathbf{R}_{\chi,z}=
    \begin{bmatrix}
	\cos\chi&\sin\chi&0\\
    -\sin\chi&\cos\chi&0\\
	   0&0&1\\
    \end{bmatrix}\!\!,
    \mathbf{R}_{-\theta_{\mathcal{R}j},y}=
    \begin{bmatrix}
	\cos\theta_{\mathcal{R}j}&0&\sin\theta_{\mathcal{R}j}\\
	0&1&0\\
	-\sin\theta_{\mathcal{R}j}&0&\cos\theta_{\mathcal{R}j}\\
    \end{bmatrix}\!\!,\\
    &\mathbf{R}_{\psi_i,z}=
    \begin{bmatrix}
	\cos(\psi_i+n\frac{\pi}{2})&\sin(\psi_i+n\frac{\pi}{2})&0\\
	-\sin(\psi_i+n\frac{\pi}{2})&\cos(\psi_i+n\frac{\pi}{2})&0\\
	0&0&1\\
    \end{bmatrix}\\
    \end{aligned}
        \label{rot_mat}
\end{equation}}
\normalsize
The coordinate transformation from the inertial frame to the blade-fixed frame of the blade is given by,
\small
\begin{equation}
  \left[ x_{3\mathcal{R}j}\,\, y_{3\mathcal{R}j}\,\, z_{3\mathcal{R}j} \right]^T = \mathbf{R}_{-\theta_{\mathcal{R}j},y}\mathbf{R}_{\psi_i,z}\mathbf{R}_{\xi,x} \mathbf{R}_{\beta,y}\mathbf{R}_{\chi,z}\left[ X\,\, Y\,\, Z \right]^T
  \label{eq:rot_seq}
\end{equation}
\normalsize
The autogyro system of Fig.~\ref{fig:gc} comprises a frame, four hubs at $A$,$B$,$D$, and $E$, and 16 blades. The kinetic energy $T$ and the potential energy $V$ associated with each component are obtained to formulate the Lagrangian, i.e., $L=T-V$. The equations of motion governing the tethered system are derived as,
\begin{equation}
    \frac{d}{dt}\left(\frac{\partial L}{\partial \dot{q}_i}\right) - \frac{\partial L}{\partial q_i} = Q_{gi}
    \label{kinematic}
\end{equation} 
where $i=1,\,2,\ldots 26$ and $q_i$ is the i\textsuperscript{th} generalized coordinate, see Eq.~\eqref{gen_coord}. Here, $Q_{gi}$ denotes the generalized forces and moments induced by aerodynamics and tether tension, the formulation of which is primarily adopted from \cite{noboni2025}. In \cite{noboni2025}, only longitudinal rotors were modeled. The present work extends the equations of \cite{noboni2025} to accommodate four rotors using the full 3-D rotational sequence defined in Eq.~\eqref{eq:rot_seq}. In this extended formulation, the aerodynamic forces and moments, as well as tether tension, are adapted to reflect the modified geometry, additional degrees of freedom, and rotor interactions. The following two paragraphs describe the resulting modifications.
\begin{figure}[htbp]
	\centering
    \includegraphics[width=0.43\textwidth]{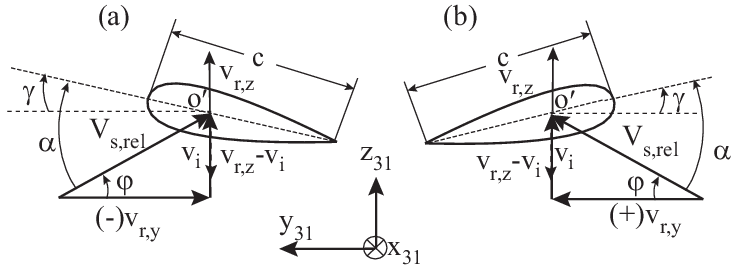}\
	\caption{Relative wind velocity at a blade element of: (a) CCW rotating blade, (b) CW rotating blade.}
	\label{fig:wind_vel}
\end{figure}

\noindent \textbf{Aerodynamic Model}: In a quad-rotor configuration, adjacent rotors rotate in opposite directions to cancel the net torque on the airframe and maintain yaw stability. To ensure that all rotors generate upward lift, the blades of adjacent rotors are geometrically mirrored. Figure~\ref{fig:wind_vel} illustrates the blade elements of adjacent rotors. The relative wind velocity experienced by an element at point $o'$ is obtained as follows:
\small
\begin{equation}
    \begin{gathered}
        \boldsymbol {v_{w/o'}}=\left(\mathbf{R}_{-\theta_{\mathcal{R}j},y}\mathbf{R}_{\psi_i,z}\mathbf{R}_{\xi,x} \mathbf{R}_{\beta,y}\mathbf{R}_{\chi,z} 
    \begin{bmatrix}
	V_w\\
    0\\ 
	0 
    \end{bmatrix}\right)^T \begin{bmatrix}
	\boldsymbol{i_{3\mathcal{R}j}}\\
     \boldsymbol{j_{3\mathcal{R}j}}\\ 
	\boldsymbol{k_{3\mathcal{R}j}}\\ 
    \end{bmatrix}-\boldsymbol{\dot{r}_{b}}\\=v_{r,x}\boldsymbol{i_{3\mathcal{R}j}}+v_{r,y}\boldsymbol{j_{3\mathcal{R}j}}+v_{r,z}\boldsymbol{k_{3\mathcal{R}j}}
        \label{rel_wind}
    \end{gathered}
\end{equation}
\normalsize
In Eq.~\eqref{rel_wind}, $\boldsymbol{\dot{r}_b}$ denotes the velocity of the blade element, the formulation of which is detailed in \cite{noboni2025} and modified using the rotational sequence given in Eq.~\eqref{eq:rot_seq}. For the mirrored adjacent rotors, the blade geometry is accounted for by reversing the sign of the y-component of the relative wind speed. ,  i.e., $v_{r,y}$ in Fig.~\ref{fig:wind_vel}. This changes the local inflow angle $\phi$, which consequently changes the angle of attack at each blade element. Finally, the aerodynamic forces and moments are computed using BEMT \citep{Gessow52_b}, with each blade discretized into 10 spanwise elements.

%A quad-rotor configuration requires adjacent rotors to spin in the opposite directions to achieve yaw stability by canceling net torque on the airframe. The aerodynamic effect of the mirrored blade is captured in this model by reversing the direction of the y component of relative wind velocity vector, i.e., $v_{r,y}$ in Fig.~\ref{fig:wind_vel}.The chord line and blade pitch remain defined with respect to the local coordinate system, which consequently modifies the local angle of attack, $\alpha$, at each blade element. The Blade Element Momentum theory \citep{Gessow52_b} is then employed to evaluate aerodynamic forces and moments, with the blades discretized into 10 spanwise elements.
% is then used to compute the aerodynamic forces and moments, with each blade divided into 10 spanwise elements.

\begin{figure}[hpbt]

	\centering
	\includegraphics[scale=0.42]{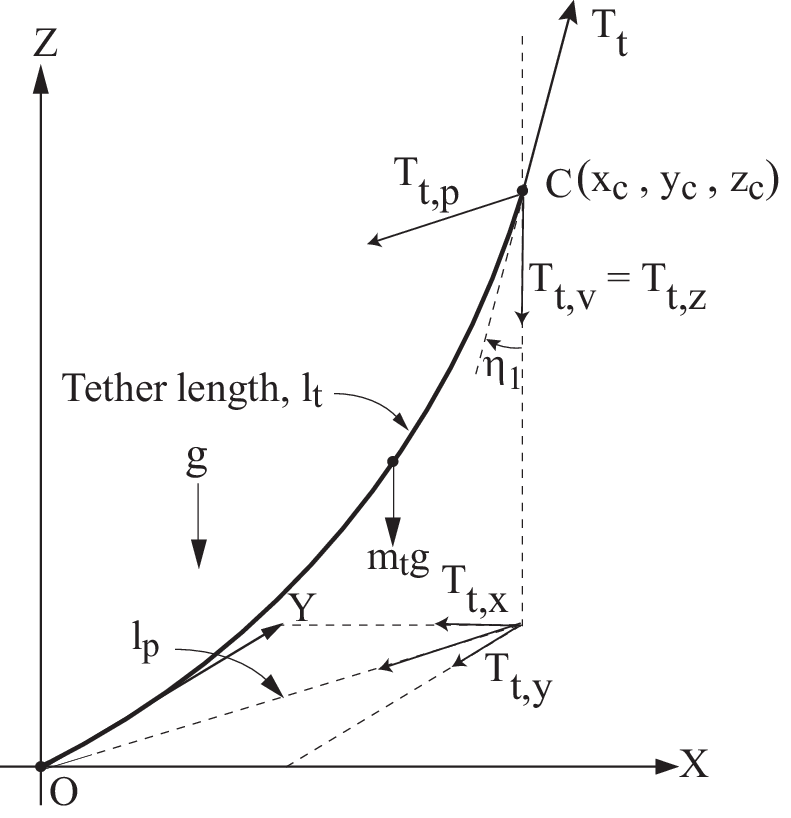}
	\caption{Static force on a catenary tether in a three-dimensional configuration.}
	\label{fig_tether}
\end{figure}

\noindent \textbf{Tether Model}: 
The planar and vertical components of the tether tension, denoted by $T_{t,p}$ and $T_{t,v}$ in Fig.~\ref{fig_tether}, are formulated by extending the static-catenary-based tether model detailed in \cite{noboni2025} to three spatial dimensions. %To mitigate numerical instabilities associated with the taut-tether, tether compliance is incorporated through an added stiffness term \citep{masciola2013implementation}.
In the resulting catenary model, $T_{t,p}$ is evaluated using the projected tether length, i.e., $l_p$, in the X-Y plane and is assumed to act radially inward toward the anchor point $O$. The modified catenary equations are given by,
\begin{equation}
\begin{aligned}
l_p &= \left(\frac{T_{t,p}}{W_t}\right)
\Bigg[
\sinh^{-1}\!\left(\frac{T_{t,v}}{T_{t,p}}\right)
\quad - \sinh^{-1}\!\left(\frac{T_{t,v}-W_t l_t}{T_{t,p}}\right)
\Bigg]\\
&+ \frac{T_{t,p} l_t}{EA}= \sqrt{x_c^2 + y_c^2} , \\[6pt]
z_c &= \left(\frac{T_{t,p}}{W_t}\right)
\Bigg[
\sqrt{1+\left(\frac{T_{t,v}}{T_{t,p}}\right)^2}
- \sqrt{1+\left(\frac{T_{t,v}-W_t l_t}{T_{t,p}}\right)^2}
\Bigg] \\
&\quad + \frac{1}{EA}
\left(T_{t,v} l_t - \frac{W_t l_t^2}{2}\right)
\end{aligned}
\label{tethereq34}
\end{equation}
where $W_t=m_tg=\sigma_t l_t g$, $\sigma_t$ is mass per unit length of tether, and $EA$ is the tether's axial stiffness.
Equation~\eqref{tethereq34} is solved numerically to obtain $T_{t,p}$ and $T_{t,v}$ using position states $x_c$, $y_c$ and $z_c$. The  planar component of tether tension, $T_{t,p}$, is then resolved into X and Y directions as follows,
\begin{equation}
      \quad T_{t,x} = T_t \sin{\eta_1} \frac{x_c}{\sqrt{x_c^2+y_c^2}}, \quad T_{t,y} = T_t \sin{\eta_1} \frac{y_c}{\sqrt{x_c^2+y_c^2}} \\
    \label{cat_thrust}
\end{equation}
where, $\boldsymbol{T_t} = T_{t,x} \boldsymbol{I} + T_{t,y} \boldsymbol{I} +T_{t,z} \boldsymbol{K}$ and  $T_t$ is the magnitude of tether tension, and $\eta_1$ is the tether angle at point C as shown in Fig.~\ref{fig_tether}.
Detailed formulations of generalized forces and torques are given in \cite{noboni2025}. Equation~\eqref{kinematic} is alternatively expressed as,
\begin{equation}
    \mathbf{A}\ddot{q}_i+\mathbf{B}=\mathbf{Q_{gi}} \quad \Rightarrow \quad \ddot{q}_i=\mathbf{A}^{-1}(\mathbf{Q_{gi}}-\mathbf{B})
    \label{kin_mat}
\end{equation} 
where, $\mathbf{A}$ is a $26 \times 26$ matrix and a function of $q_i$ whereas $\mathbf{B}$ is a $26 \times 1$ matrix and a function both $q_i$ and $\dot{q}_i$. Equation~\eqref{kin_mat} is solved to obtain $\ddot{q}_i$ and numerically integrated twice to obtain $\dot{q}_i$ and ${q}_i$.
%In modeling the tether tension, denoted by $T_t$ in Fig.~\ref{fig_tether}, static catenary mechanics \cite{Rimkus2013StabilityAO} is used by neglecting the aerodynamic loads on the tether. The projection of $T_t$ in the X-Y plane and its vertical component, namely $T_{t,p}$ and $T_{t,v}$ in Fig.~\ref{fig_tether} respectively, are obtained by extending the formulations detailed in \cite{noboni2025} to 3 spatial dimensions. In the modified catenary equations, $T_{t,p}$ is calculated using the projection of tether length, i.e., $l_p$, in the X-Y plane, with the underlying assumption that $T_{t,p}$ always acts radially inward toward the anchor point O. The modified catenary equations are, 
\section{Problem Definition and Control Design}
\label{CL}
\vspace{-0.1in}
The coupled dynamics of the system, arising from interactions among aerodynamics, tether force, and rigid-body motions, make the system inherently unstable without active control, preventing convergence to a steady configuration in Cartesian space. Consequently, feedback control is required to regulate the autogyro’s attitude and to characterize physically meaningful equilibrium configurations.
\begin{figure}[hpbt]
\vspace{-0.1in}
	\centering
	\includegraphics[width=0.48\textwidth]{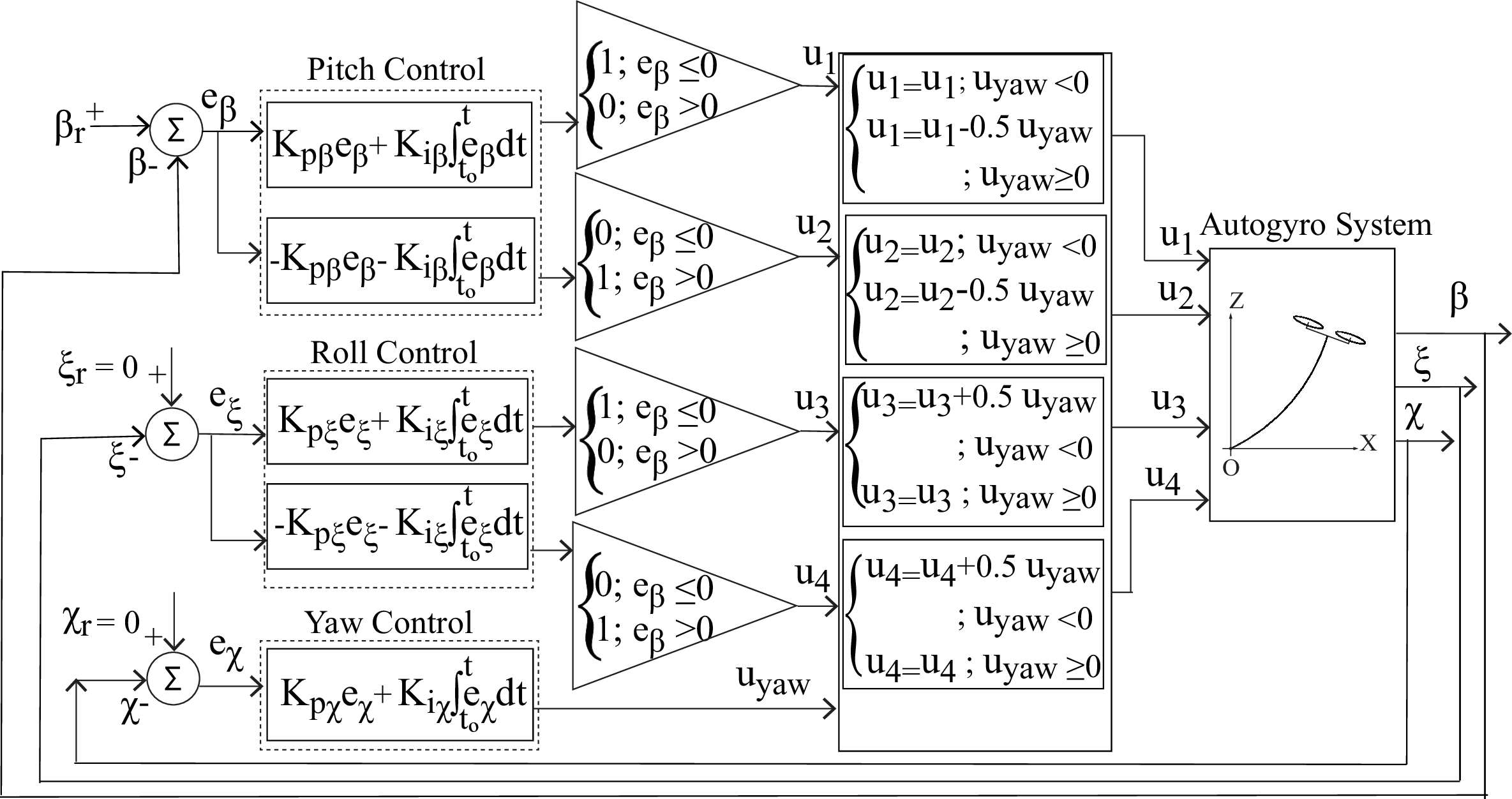}
    \vspace{-0.1in}
	\caption{\centering Schematic of proposed attitude controller.}
    \label{fig:P_schm}

\end{figure}
\begin{figure*}[tpb]
	\begin{center}
	\includegraphics[width=\textwidth]{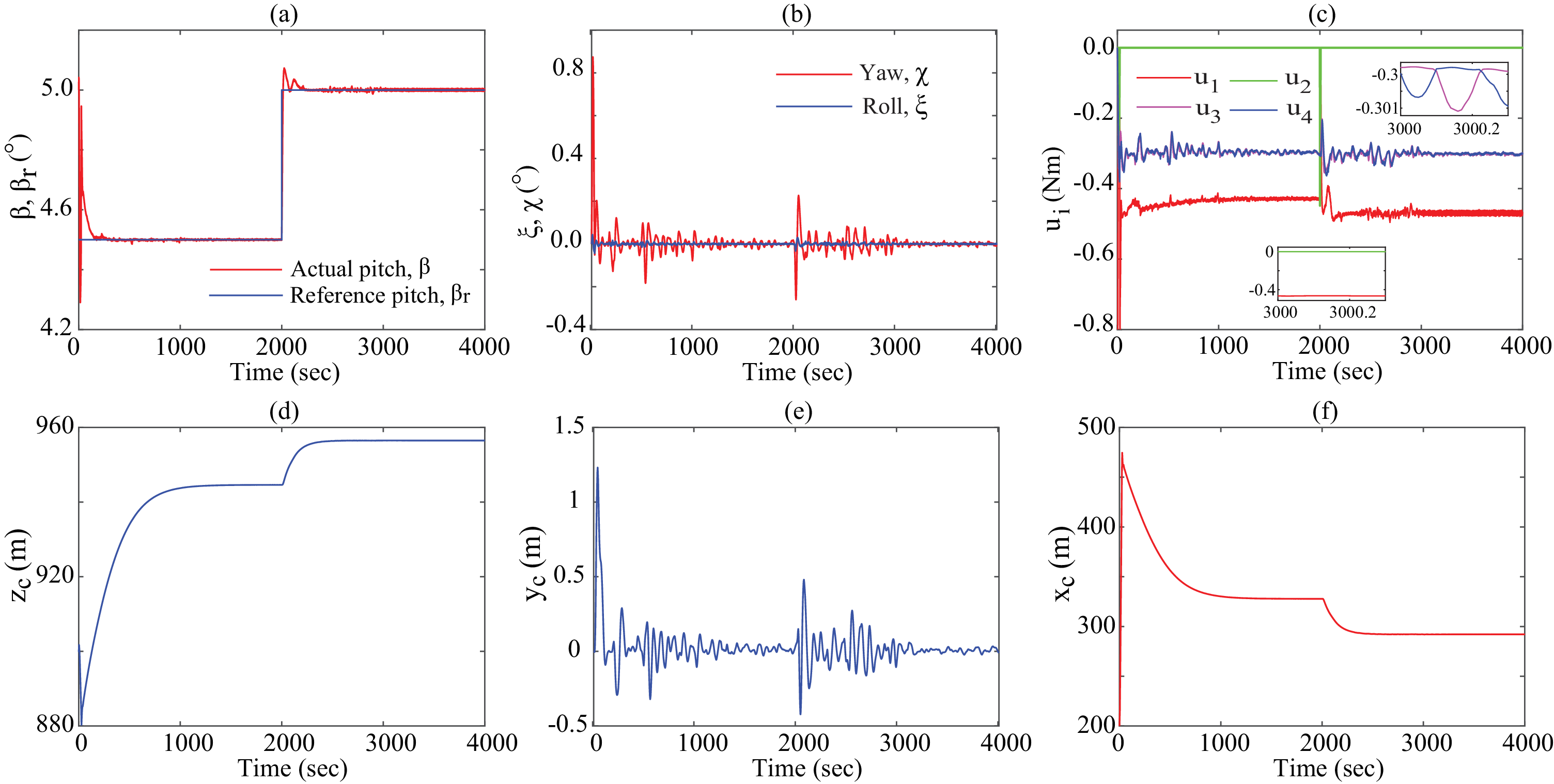}
    \vspace{-0.2in} 
        \caption{Controller performance in $V_w$=10~m/s: (a) Reference and actual pitch angle; (b) Roll and yaw angles; (c) Braking torques as control inputs; (d) Altitude; (e)Lateral position; (f) Longitudinal position.}
    \label{fig:RS1}
    \end{center}   
\end{figure*}
%In this paper, we therefore address the problem of orientation stabilization of the tethered autogyro without actuating the tether length. A feedback control algorithm based on a differential rotor braking method \citep{noboni2025} for attitude regulation of the tethered autogyro is developed. In this method, braking torques are applied regeneratively to produce a thrust difference between opposing rotors. The resulting differential thrust modulates the roll, pitch, and yaw angle of the system. Building upon our previous work \citep{noboni2025}, where pitch modulation was achieved through differential braking of two autorotating longitudinal rotors, the present control design extends the concept to a four-rotor configuration by braking the autorotating lateral rotors. It relaxes the assumption of \cite{noboni2025} that the roll and yaw motions are already being controlled by lateral rotors. These braking torques, denoted by $u_i$ where $i=1,\ldots 4$, are chosen to be control inputs in this study and are incorporated in the dynamics through equations of motion associated with the azimuth angle $\psi_i$ in Eq.~\eqref{kinematic} as follows:
In this paper, we therefore address the problem of orientation stabilization of the tethered autogyro without actuating the tether length. A feedback control algorithm based on differential rotor braking \citep{noboni2025} is developed for attitude regulation. In this approach, regenerative braking torques are applied to produce thrust differences between opposing rotors, thereby modulating the roll, pitch, and yaw angles of the system. Building on our previous work \citep{noboni2025}, where pitch modulation was achieved through differential braking of two longitudinal rotors, the present design extends the concept to a four-rotor configuration by additionally braking the autorotating lateral rotors. These braking torques, denoted by $u_i$ where $i=1,\ldots 4$, are chosen to be control inputs in this study and are incorporated in the dynamics through equations of motion associated with the azimuth angle $\psi_i$ in Eq.~\eqref{kinematic} as follows:
\begin{equation}
    \frac{d}{dt}\left(\frac{\partial L}{\partial \dot{\psi}_i}\right) - \frac{\partial L}{\partial \psi_i} = Q_{\psi_i}+\sigma_i u_i 
 \label{EOM_con}
\end{equation}
where $\sigma_i = +1$ for longitudinal rotors and $\sigma_i = -1$ for lateral rotors, representing their respective spin directions. The torques $u_i\leq0 $ act as braking torques in autorotation mode, whereas $u_i>0 $ means power is being added to the system and can be utilized to maintain altitude in low-wind conditions. The proposed control strategy is formulated as a multi-variable control problem with 4 inputs and 3 outputs. The inputs are braking torques ($u_1$, $u_2$, $u_3$, $u_4$) in each rotor, and the outputs are the Euler angles $\beta$, $\xi$, and $\chi$.

Figure~\ref{fig:P_schm} shows the feedback control structure modulating the attitude of the system. The control loop employs three independent PI controllers that track the attitude references $(\beta_r, \xi_r = 0, \chi_r = 0)$ by generating the braking torques applied to the rotors. Specifically, the torque pair $(u_1,u_2)$ applied to the longitudinal rotors regulates the pitch angle, $\beta$ while $(u_3,u_4)$ applied to the lateral rotors regulates roll angle, $\xi$. Yaw control is achieved through a dedicated PI controller that generates a yaw correction torque $u_{yaw}$. This torque is then distributed across the four rotors according to a reaction-based allocation strategy, determined by the yaw direction. During positive yawing of the system, i.e., $\chi >0$, a clockwise reaction torque is produced. To mitigate this, braking on the lateral rotors is increased, reducing their relative speeds with respect to the longitudinal rotors. Conversely, during negative yawing $\chi < 0$, a counterclockwise reaction torque is generated, and braking on the longitudinal rotors is increased to reduce their relative speed with respect to the lateral rotors. This reaction-based allocation of braking torques enables effective yaw stabilization while maintaining the constraint of braking-only actuation.
The resulting control law is,
\small
\begin{equation}
\begin{gathered}
% --- Pitch control ---
u_1=\left\{
\begin{array}{ll}
K_{p\beta}(\beta_r - \beta) + K_{i\beta}\int_{t_0}^{t} (\beta_r - \beta)\, dt, & \quad \beta > \beta_r \\
0, & \quad \beta \le \beta_r\
\end{array}
\right. \\[1.0em]
u_2=\left\{
\begin{array}{ll}
0, & \quad \beta \ge \beta_r \\
-K_{p\beta}(\beta_r - \beta) - K_{i\beta}\int_{t_0}^{t} (\beta_r - \beta)\, dt, & \quad \beta < \beta_r
\end{array}
\right. \\[1.0em]
% --- Roll control ---
u_3=\left\{
\begin{array}{ll}
K_{p\xi}(\xi_r - \xi) + K_{i\xi}\int_{t_0}^{t} (\xi_r - \xi)\, dt, & \quad \xi > \xi_r \\
0, & \quad \xi \le \xi_r
\end{array}
\right. \\[1.0em]
u_4=\left\{
\begin{array}{ll}
0, & \quad \xi \ge \xi_r \\
- K_{p\xi}(\xi_r - \xi) - K_{i\xi}\int_{t_0}^{t} (\xi_r - \xi)\, dt, & \quad \xi < \xi_r
\end{array}
\right. \\[1.0em]
% --- Yaw control (always active) ---
u_{yaw} = K_{p\chi}(\chi_r - \chi) + K_{i\chi} \int_{t_0}^{t} (\chi_r - \chi)\, dt \\ \text{where,} \quad \text{If } u_{yaw} < 0: \quad \begin{cases}
u_1 = u_1 \\
u_2 = u_2 \\
u_3 = u_3 + \tfrac{1}{2} u_{yaw} \\
u_4 = u_4 + \tfrac{1}{2} u_{yaw} \quad
\end{cases} \\
\quad \text{\& \; If } u_{yaw} \ge 0: \quad \begin{cases}
u_1 = u_1 - \tfrac{1}{2} u_{yaw} \\
u_2 = u_2 - \tfrac{1}{2} u_{yaw} \\
u_3 = u_3\\
u_4 = u_4
\end{cases}
\end{gathered}
\label{Eq:PID_control}
\end{equation}
\normalsize
where, $K_{p_\beta}$, $K_{p_\xi}$, $K_{p_\chi}$ and $K_{i_\beta}$, $K_{i_\xi}$, $K_{i_\chi}$ are the proportional and integral gains of the PI controllers. Equation \eqref{Eq:PID_control} is formulated with the constraint $u_i\le0$, thereby ensuring that the actuators provide only braking action on the rotors. The magnitude of $u_i$ is restricted to small values to prevent rapid variations in the Euler angles.
%\vspace{-0.1in}
\section{Simulation Results}
\begin{figure}[htbp]
	\centering
	\includegraphics[width=0.48\textwidth]{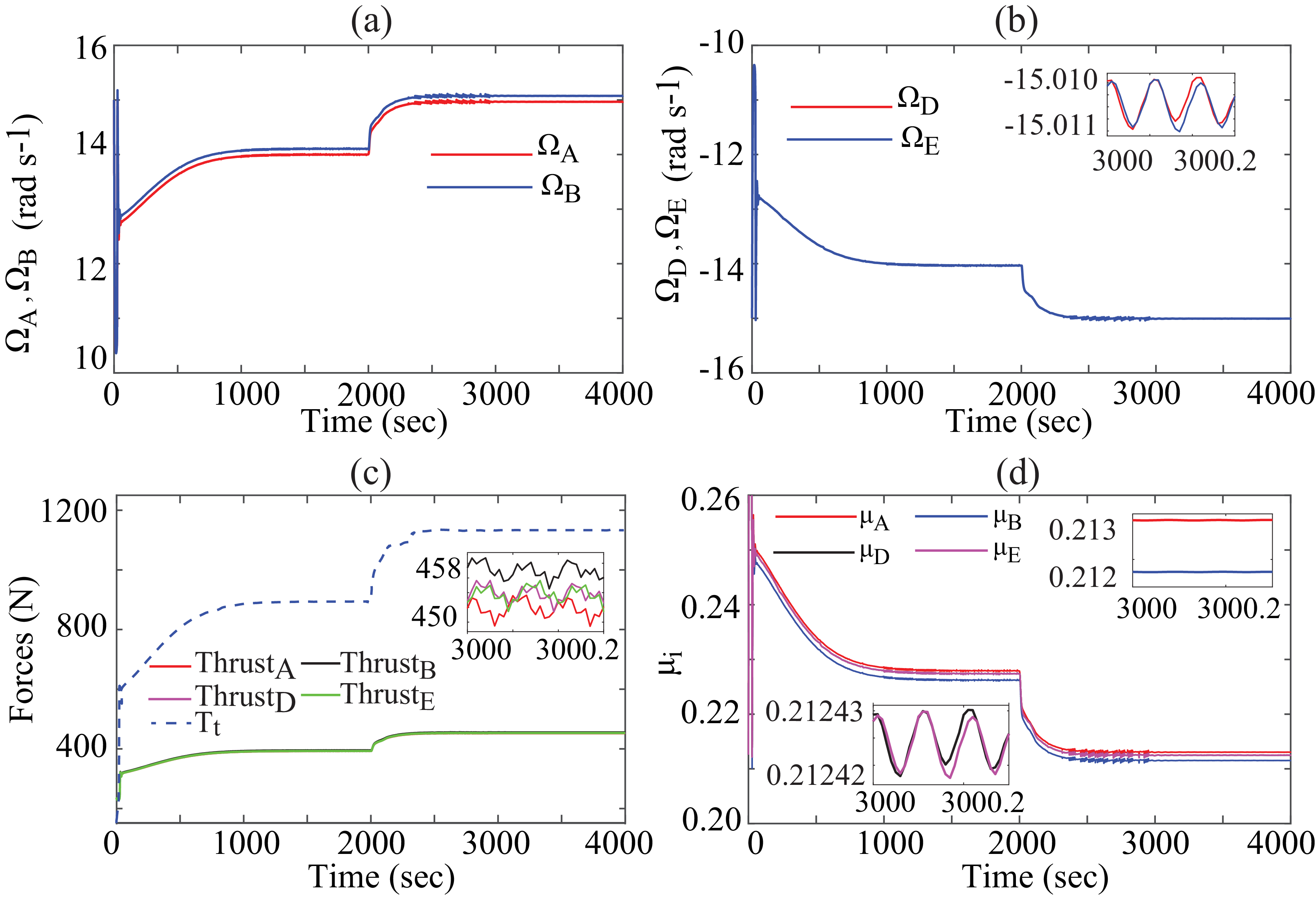}
        \vspace{-0.1in}
	    \caption{Controller performance in $V_w$=10~m/s: (a) Longitudinal rotor speeds; (b) Lateral rotor speeds; (c) Tether tension and thrust forces; (d) Tip-speed ratios}
        \vspace{-0.1in}
	\label{fig:Rs2_states}
\end{figure}

\vspace{-0.1in}
The proposed attitude controller in Sec.~\ref{CL} is implemented using aerodynamic parameters from \cite{10155811, noboni2025} and updated physical parameters given in Table~\ref{TB:pres_tb} to account for the heavier quadrotor configuration. The equations of motion in Eq.~\eqref{kin_mat} are solved numerically by providing initial values.
 \begin{table}[hbt!]
\begin{center}
\caption{Physical parameters of the system} \label{TB:pres_tb}
\begin{tabular}{cccc}
\hline
Parameter & Numerical Value (with units)\\  
\hline
\hline
$m_f, m_h, m_b$ & $27.2~kg, 1~kg, 2.5418~kg$\\
$l, r_h, d, R$ & $11.754~m, 0.0762~m, 0.03048~m, 3.0480~m$  \\
$\sigma_t$ & $0.022~kg/m$ \\
$EA$ & $4.17\times10^6~N$ \\
\hline
\end{tabular}
\end{center}
%\vspace{-0.3in}
\end{table}
The performance of the proposed attitude controller is illustrated in Fig.~\ref{fig:RS1} and Fig.~\ref{fig:Rs2_states} for a fixed tether length of $1000$~m and in uniform wind speed, $V_w$=10~m/s. The magnitudes of controller gains $K_{p\beta}, \, K_{p\xi}=100$, $K_{p\chi}=25$ and $K_{i\beta}, \, K_{i\xi},\, K_{i\chi}=2$ are used in each PI controller. The control objective is to regulate the pitch angle $\beta$ to a commanded reference $\beta_r$, while the roll angle, $\xi$, and yaw angle, $\chi$, are driven to zero-reference setpoints to maintain lateral and yaw stability.
\vspace{-0.1in} 

Figure~\ref{fig:RS1}(a) shows that $\beta$ converges to $\beta_r$ at $4.5^\circ$ and $5^\circ$ with negligible error.
During the transition in $\beta$, the yaw angle $\chi$ and roll angle $\xi$ exhibit small transient deviations from zero, as shown in Fig.~\ref{fig:RS1}(b). Specifically, $\chi$ reaches a peak deviation of approximately $0.2^\circ$, while $\xi$ shows a smaller deviation of about $0.03^\circ$. Both angles subsequently converge to $0^\circ$ due to the braking of the rotors. The control inputs are the braking torques, $u_i$, applied to the rotors, as shown in Fig.~\ref{fig:RS1}(c), and are constrained within the range $-1~\text{Nm} \le u_i \le 0~\text{Nm}$ to avoid drastic changes in the Euler angles. The braking torques are not identical as seen in the zoomed view, which highlights small but persistent differences among $u_i$, generating differential rotor speeds where $\Omega_A \neq \Omega_B$ and $\Omega_D \neq \Omega_E$. This imbalance produces differential thrust in alternate rotors, modulating the attitude angles. The control inputs $u_i$ in Fig.~\ref{fig:RS1}(c) show that nonzero braking is required even at equilibrium to maintain the desired orientation. In particular, continuous braking of rotor 1 maintains the equilibrium $\beta$, while that of the lateral rotors compensates for the system’s positive yawing tendency, i.e., $\chi \geq 0$. Figure~\ref{fig:RS1}(e) shows that the lateral drift, $y_c$, induced by the coupling during pitch transitions, converges to zero as $\xi \to 0$, implying that lateral drift can be controlled with roll actuation. The altitude $z_c$ and horizontal position $x_c$ in Figs.~\ref{fig:RS1}(d) and (f), respectively, reflect the shift in equilibrium corresponding to the change in $\beta$. As $\beta$ increases, the autogyro ascends to a higher $z_c$ while $x_c$ decreases, which is consistent with the results reported in \cite{noboni2025}.
\begin{figure}[htbp]
	\centering
	\includegraphics[trim = 420 40 450 50, clip, scale=0.32]{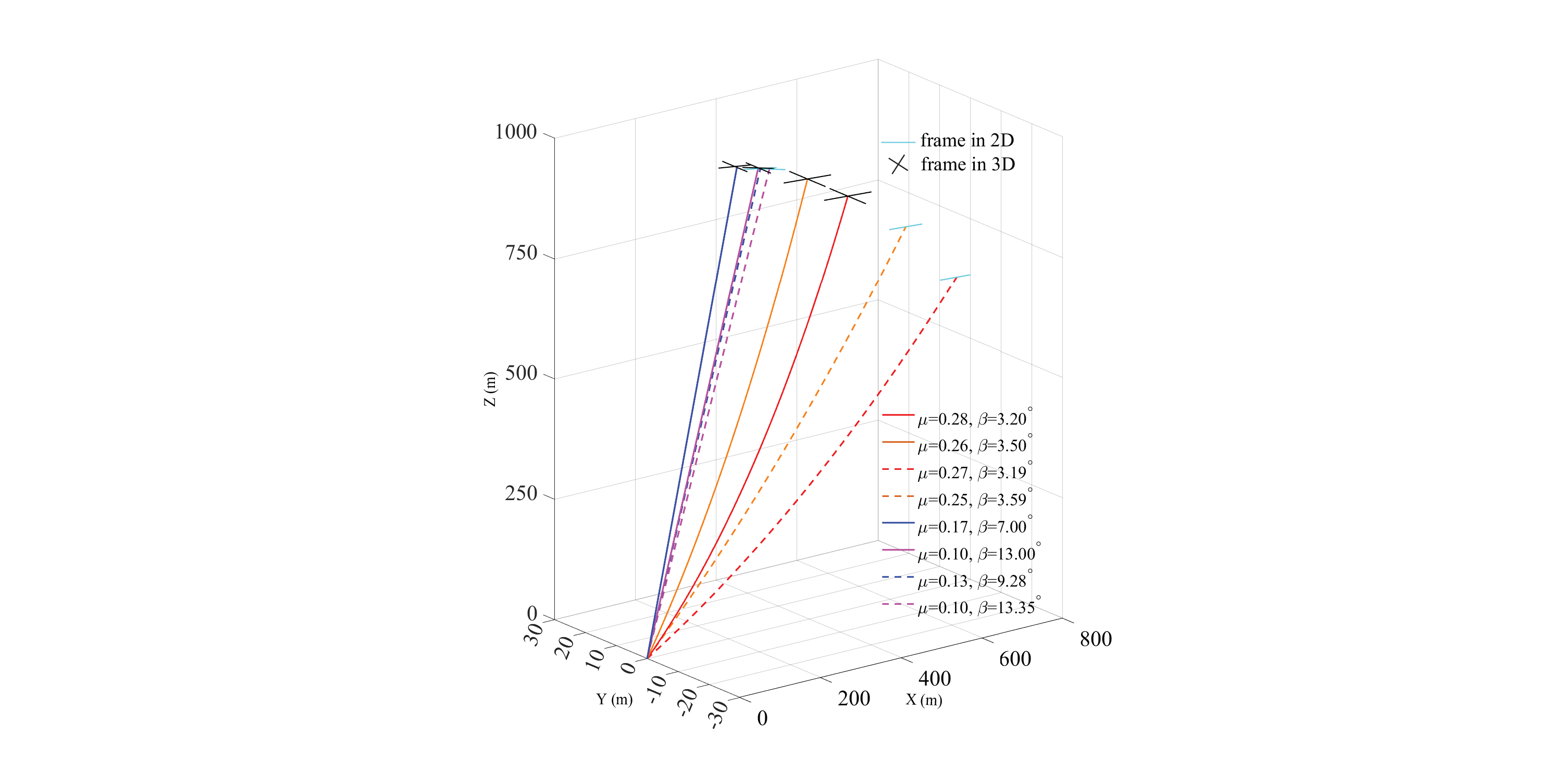}
        \vspace{-0.1in}
	    \caption{Equilibrium tether profiles of the autogyro with ${l_t}$= 1000~m in ${V_w}$= 10~m/s}
	\label{fig:eqbvis}
\end{figure}

\vspace{-0.1in}
Figures~\ref{fig:Rs2_states}(a) and (b) present the angular speeds of the longitudinal rotors ($\Omega_A, \Omega_B$) and lateral rotors ($\Omega_D, \Omega_E$), respectively. Following the step change in $\beta_r$ at $t = 2000$~s, the magnitudes of all rotor speeds increase. A small but persistent difference in rotor speeds between opposing rotors is observed, generating the required differential thrust to regulate attitude. %The zoomed view of the lateral rotor speeds in Fig.~\ref{fig:Rs2_states}(b) shows a small difference between opposing rotors required to maintain $\xi$ and $\chi$ near zero.The zoomed plot of Fig.~\ref{fig:Rs2_states}(c) indicates that the thrust forces of the four rotors do not become identical even after $\beta \rightarrow \beta_r$ and $\xi, \chi \rightarrow 0$. This persistent differential thrust confirms the effectiveness of the controller in maintaining the desired operating condition. 
The zoomed plot in Fig.~\ref{fig:Rs2_states}(c) shows that persistent differential thrust remains among the four rotors due to the continuous braking observed in Fig.~\ref{fig:RS1}(c), even after $\beta \to \beta_r$ and $\xi,\chi \to 0$. Figure~\ref{fig:Rs2_states}(c) also exhibits that the tether tension, $T_t$, increases with $\beta$ as the altitude $z_c$ rises. This indicates that the reference pitch angle $\beta_r$ must be selected carefully to prevent excessive tether tension during flight. The tip speed ratio $\mu$, i.e., the ratio of the wind speed parallel to the rotor disc to the speed of the rotor blade tip, is shown in Fig.~\ref{fig:Rs2_states}(d). As $\beta$ increases from $4.5^\circ$ to $5^\circ$, the value of $\mu$ decreases from approximately $0.23$ to $0.21$. These values remain well within the theoretical validity range of momentum theory, that is $0.1 < \mu < 0.5$ \citep{wheatley1935aerodynamic,mcconnell2022equilibrium}. The zoomed view further shows negligible differences in the tip-speed ratios due to small rotor-speed variations under differential braking.
\vspace{-0.1in}
\subsection{Characteristics of Equilibria}
\label{EQB}
\vspace{-0.1in}
The equilibrium condition is characterized by computing the steady-state response of the system for varying pitch angles $\beta$ in the presence of the attitude controller while maintaining the lateral drift, i.e., $y_c = 0$. 
 %\vspace{-0.1in}
\begin{figure}[htbp]
	\begin{center}
	\includegraphics[width=0.46\textwidth]{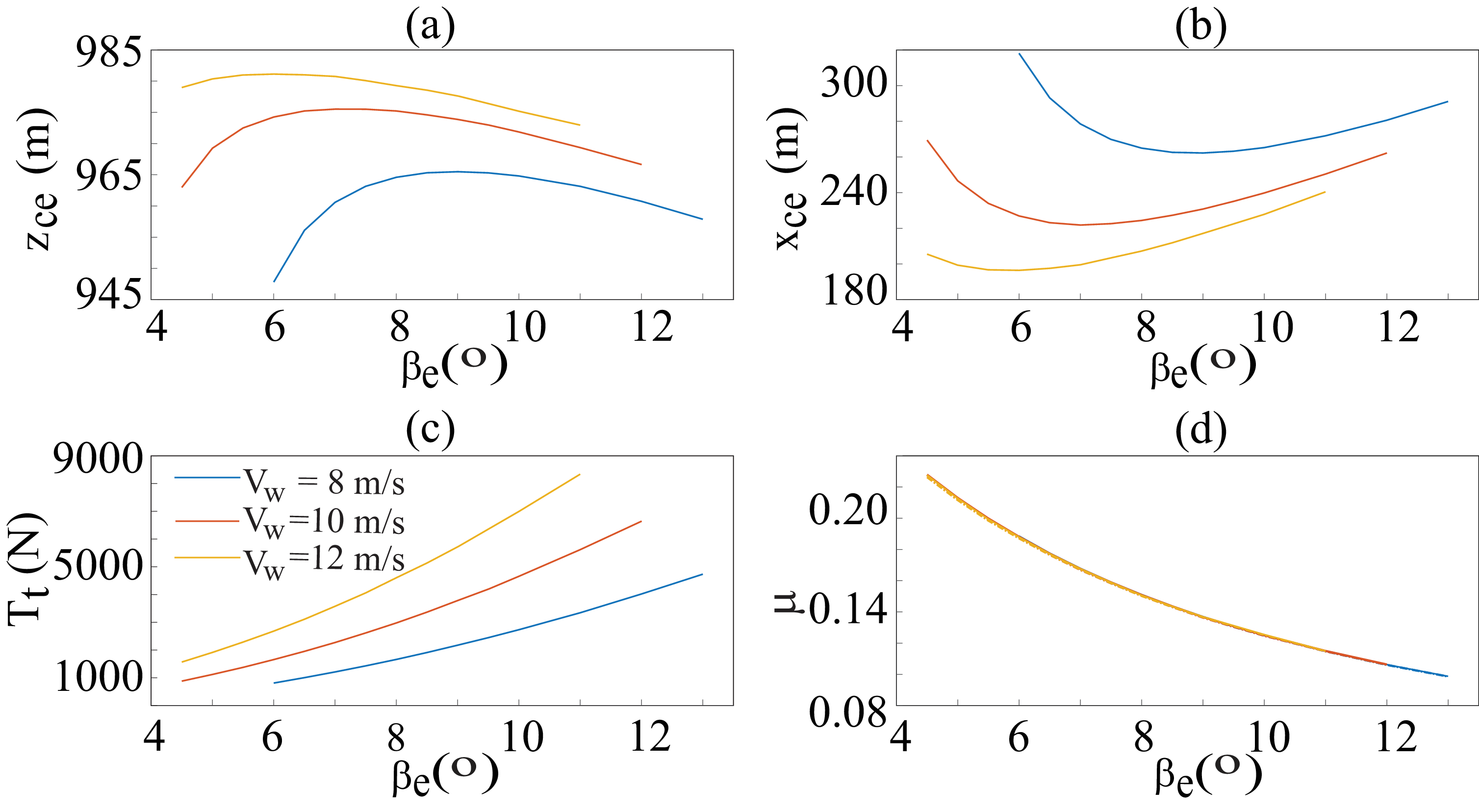}
            \vspace{-0.1in}
	    \caption{Equilibrium characteristics with ${l_t}$= 1000m in varying $V_w$: (a) Altitude; (b) Horizontal drift; (c) Tether tension; (d) Tip speed ratio.}
	\label{fig:beta_muTxz_vwvary}
    \end{center}
\end{figure}
 \vspace{-0.1in}
 
Figure~\ref{fig:eqbvis} compares the equilibrium tether profiles of the quadrotor-based autogyro with $l_t=1000$~m and in a uniform wind speed of $10$~m/s, obtained from the hybrid model developed in \cite{noboni2025} and the 3D model developed in this paper, represented by solid and dashed lines, respectively. A scaled-up frame is used for clarity in Fig~\ref{fig:eqbvis}. At lower $\beta$, the reduced lift leads to lower altitude $z_c$, greater horizontal drift $x_c$, and lower tether tension. As $\beta$ increases, altitude initially rises and reaches a maximum at a certain pitch angle, after which further increase in $\beta$ causes $z_c$ to decrease. In the 3D model, the maximum altitude occurs near $\beta=7^\circ$, corresponding to $\mu=0.17$, whereas in the hybrid model, the maximum $z_c$ occurs at the higher pitch angle of $\beta=9.28^\circ$. This shift indicates an increase in total lift capacity due to the inclusion of the lateral rotor pair. Beyond this point, drag effects become dominant, causing $z_c$ to drop and resulting in a taut tether.
\vspace{-0.2in}
\begin{figure}[htbp]
	\begin{center}
	\includegraphics[width=0.46\textwidth]{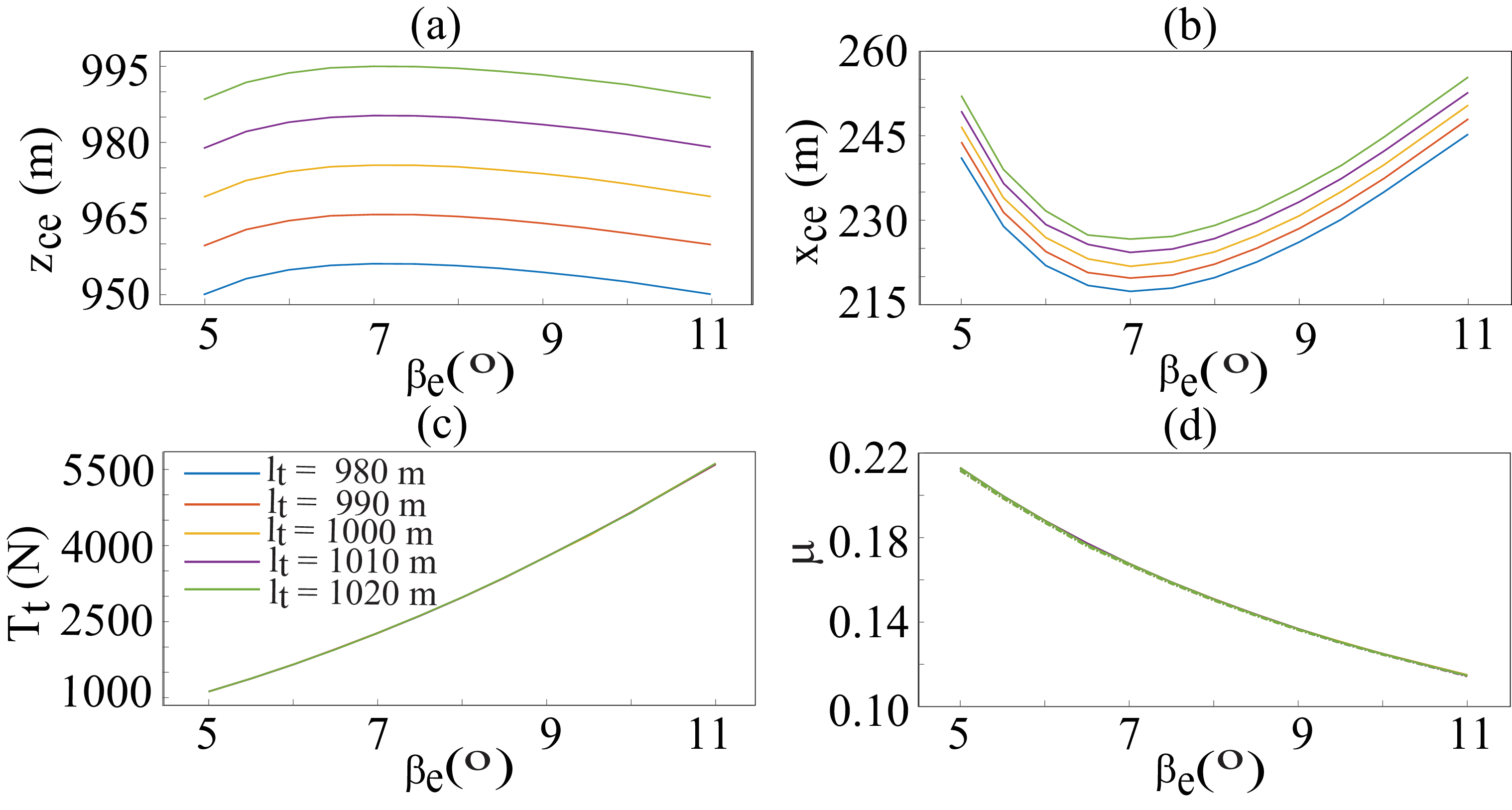}
            \vspace{-0.1in}
	    \caption{Equilibrium characteristics with varying $l_t$ in $V_w$= 10~m/s: (a) Altitude; (b) Horizontal drift; (c) Tether tension; (d) Tip speed ratio.}
	\label{fig:beta_muTxz_ltvary}
    \end{center}
    \vspace{-0.2in}
\end{figure}

Figures~\ref{fig:beta_muTxz_vwvary} and~\ref{fig:beta_muTxz_ltvary} show the equilibrium space with $\beta$ for varying $V_w$ at fixed tether length $l_t=1000~\mathrm{m}$, and for varying $l_t$ at fixed wind speed $V_w=10~\mathrm{m/s}$, respectively. Consistent with the equilibrium trend in Fig.~\ref{fig:eqbvis}, Figs.~\ref{fig:beta_muTxz_vwvary}(a) and~\ref{fig:beta_muTxz_ltvary}(a) show that, up to a certain pitch angle, the equilibrium altitude $z_{ce}$ increases with $\beta$, while Figs.~\ref{fig:beta_muTxz_vwvary}(b) and~\ref{fig:beta_muTxz_ltvary}(b) show the corresponding reduction in horizontal drift $x_{ce}$. Increasing $V_w$ leads to a higher $z_{ce}$ for the same $\beta$ and shifts the maximum $z_{ce}$ to a lower $\beta$, as seen in Fig.~\ref{fig:beta_muTxz_vwvary}(a). Figures~\ref{fig:beta_muTxz_ltvary}(a)-(b) shows increasing $l_t$ at constant $V_w$ shifts the $z_{ce}$ and $x_{ce}$ curves upward, allowing different equilibrium positions at the same $\beta$. The equilibrium tether tension, $T_t$, increases with $\beta$ and becomes larger at higher $V_w$, but remains nearly unchanged with varying $l_t$, as shown in Figs.~\ref{fig:beta_muTxz_vwvary}(c) and~\ref{fig:beta_muTxz_ltvary}(c), respectively. Finally, Figs.~\ref{fig:beta_muTxz_vwvary}(d) and~\ref{fig:beta_muTxz_ltvary}(d) show that, for a given $\beta$, $\mu$ remains nearly unchanged with variations in $V_w$ and $l_t$. These trends are consistent with prior results reported in \cite{mcconnell2022equilibrium,10155811,noboni2025}, thereby validating the developed 3D model as a reliable basis for subsequent altitude and position tracking control design.
\vspace{-0.1in}
\section{Conclusion}
\vspace{-0.12in}
A 3D dynamic model of a tethered quadrotor autogyro with articulated rotors is developed in this paper to capture the complete rigid-body motion of the system. The agreement of the observed equilibrium characteristics with previously reported trends supports the validity of the model. Using this framework, a feedback controller based on regenerative differential rotor braking is designed for attitude regulation, and simulations demonstrate accurate pitch tracking together with stable roll and yaw regulation, while maintaining overall system stability.
%A 3D dynamic model of a tethered autogyro with articulated rotors is developed in this paper by extending the hybrid model to capture the system's complete rigid-body motion. The observed equilibrium characteristics align with previously reported trends, supporting the validity of the model. An attitude controller based on regenerative differential rotor braking is proposed for full attitude regulation, and controller implementation via simulations demonstrates accurate pitch tracking together with stable roll and yaw regulation of the system. 
%A regenerative differential rotor braking strategy is developed for full attitude regulation of a tethered multirotor autogyro using a three-dimensional dynamic model. Simulation results demonstrate accurate pitch tracking with stable roll and yaw regulation. The equilibrium characteristics under varying wind speeds and tether lengths align with previously reported trends, confirming that the model captures the essential aerodynamic–tether interactions. 
\vspace{-0.1in}
\bibliography{citation_file}             % bib file to produce the bibliography
                                                     % with bibtex (preferred)
                                                   
%\begin{thebibliography}{xx}  % you can also add the bibliography by hand

%\bibitem[Able(1956)]{Abl:56}
%B.C. Able.
%\newblock Nucleic acid content of microscope.
%\newblock \emph{Nature}, 135:\penalty0 7--9, 1956.

%\bibitem[Able et~al.(1954)Able, Tagg, and Rush]{AbTaRu:54}
%B.C. Able, R.A. Tagg, and M.~Rush.
%\newblock Enzyme-catalyzed cellular transanimations.
%\newblock In A.F. Round, editor, \emph{Advances in Enzymology}, volume~2, pages
%  125--247. Academic Press, New York, 3rd edition, 1954.

%\bibitem[Keohane(1958)]{Keo:58}
%R.~Keohane.
%\newblock \emph{Power and Interdependence: World Politics in Transitions}.
%\newblock Little, Brown \& Co., Boston, 1958.

%\bibitem[Powers(1985)]{Pow:85}
%T.~Powers.
%\newblock Is there a way out?
%\newblock \emph{Harpers}, pages 35--47, June 1985.

%\bibitem[Soukhanov(1992)]{Heritage:92}
%A.~H. Soukhanov, editor.
%\newblock \emph{{The American Heritage. Dictionary of the American Language}}.
%\newblock Houghton Mifflin Company, 1992.

%\end{thebibliography}

%\appendix
%\section{A summary of Latin grammar}    % Each appendix must have a short title.
%\section{Some Latin vocabulary}              % Sections and subsections are supported  
                                                                         % in the appendices.
\end{document}